%% file: main.tex
\documentclass[sigconf, screen, review]{acmart}

\usepackage{style}

\AtBeginDocument{%
  }

\setcopyright{none} 
\copyrightyear{2026}
\acmYear{2026}
\acmDOI{XXXXXXX.XXXXXXX}

\acmConference[MICRO 2026]{The 58th IEEE/ACM International Symposium on Microarchitecture}{October 31--November 04, 2026}{Athens, Greece}

\acmISBN{978-X-XXXX-XXXX-X/XX/XX}

\begin{document}

\title{The Turbo-Charged Mapper: Fast and Optimal Mapping for Energy-efficient and Low-latency Accelerator Design}



\author{Michael Gilbert}
\authornote{These authors contributed equally to this work.}
\email{gilbertm@mit.edu}
\affiliation{%
  \institution{MIT}
  \city{Cambridge}
  \state{MA}
  \country{USA}
}

\author{Tanner Andrulis}
\authornotemark[1]
\email{andrulis@mit.edu}
\affiliation{%
  \institution{MIT}
  \city{Cambridge}
  \state{MA}
  \country{USA}
}

\author{Vivienne Sze}
\email{sze@mit.edu}
\affiliation{%
  \institution{MIT}
  \city{Cambridge}
  \state{MA}
  \country{USA}
}

\author{Joel S. Emer}
\email{emer@csail.mit.edu}
\affiliation{%
  \institution{MIT / Nvidia}
  \city{Cambridge}
  \state{MA}
  \country{USA}
}


\input{sections/0.abstract}
\maketitle

\input{sections/1.introduction}

\input{sections/2.background}

\input{sections/3.body}

\input{sections/4.body2}

\input{sections/5.results}

\input{sections/7.conclusion}


\bibliographystyle{ACM-Reference-Format}
\bibliography{refs}

\end{document}

%% file: sections/0.abstract.tex
\begin{abstract}
The energy and latency of an accelerator running a deep neural network (DNN) depend on how the computation and data movement are scheduled in the accelerator (i.e., mapping), and picking an optimal mapping is essential to achieve high-performance accelerators. However, it is challenging to find mappings that maximize accelerator performance. The space of mappings is large, and prior works cannot guarantee finding optimal mappings because they use heuristics or metaheuristics to narrow the search space.

To address this challenge, we propose the Turbo-Charged Mapper (TCM), a fast mapper that finds optimal mappings. The key to our approach is that we define a new mapping concept called dataplacement, which, like the prior concept of dataflow, allows for clear analysis and comparison of mappings. Through it, we identify opportunities to prune redundant and suboptimal mappings, reducing search space by up to 32 orders of magnitude ($10^{37}\rightarrow10^5$).

TCM leverages these insights to perform full mapspace searches, making it the first mapper that can find optimal mappings in feasible runtime. Compared to prior mappers, TCM improves accelerator energy-delay-product by $1.2-6.5\times$ while simultaneously reducing mapping search time by $1000\times$ (5 hours $\rightarrow$ 17 seconds).
\end{abstract}

%% file: sections/1.introduction.tex


\section{Introduction}
\def\thefootnote{*}\footnotetext{These authors contributed equally to this work}\def\thefootnote{\arabic{footnote}}

Achieving \blfootnote{\label{url} TCM is available at \emph{<to be posted after peer review.>}}
low energy and latency with deep neural network (DNN) accelerators require various optimizations such as parallelization and using data reuse to reduce expensive data movement. These optimizations depend on the accelerator architecture, but also how computation and data movement are scheduled onto the accelerator (\ie the \emph{mapping}~\cite{timeloop, maestro, tileflow, set, looptree}). Various mapping optimizations have been proposed, and mapping proposals often open a wider space of mappings (\ie \emph{mapspace}~\cite{timeloop}). For example, ZigZag~\cite{zigzag} introduced the concept of \emph{uneven mappings}, which expanded the space of possible data tiles kept in each memory level.

Although a wider mapspace may offer more efficient mappings and enable better-performing accelerators, the challenge for the \emph{mapper}~\cite{timeloop}, which searches for an optimal mapping\footnote{\emph{Optimal} means that, in the defined mapspace, there is no valid (\ie within resource limits) mapping with superior objective metrics (\eg energy). We also consider redundant mappings (\ie identical metrics to an unpruned optimal mapping) to be suboptimal. While we cannot predict optimizations that will be invented in the future, our optimality guarantee applies in a superset of the mapspaces of many prior works~\cite{timeloop,zigzag,loma,sunstone,cosa,maestro,orojenesis}.}, becomes more difficult. Current mappers must search very large space of mappings, containing as many as $10^{37}$ mappings\footnote{Mapping counts in the introduction are from Section~\ref{section:pruning_rate}, and depend on workload, architecture, and mapspace}. Given the infeasibility of evaluating every mapping, prior mappers~\cite{zigzag,timeloop,set,tileflow,sunstone,loma} use heuristics (\eg only evaluate mappings that maximize buffer utilization) or metaheuristics (\eg randomly sample different mappings), which speed up mapspace search but cannot guarantee finding an optimal mapping. This lack of guarantee leads to significant practical limitations, such as significant accelerator energy-delay-product degradations (Section~\ref{sec:prior_work_comparison}) and misinforming architecture explorations (Section~\ref{sec:case_study}).

\tanner{Check FFM for optimal typo}


To address this challenge, we introduce new strategies for pruning suboptimal mappings from the mapspace. These strategies reduce search size by up to 32 orders of magnitude while guaranteeing finding an optimal mapping. These pruning strategies leverage new mapping analyses, enabled by a new concept called \emph{dataplacement}.

A mapping comprises dataplacement, which we define as the choice of which tiles are kept at a given point in time in each memory level of the accelerator, alongside \emph{tile shapes} and \emph{dataflow}. Prior work has defined \emph{tile shapes} as the size and number of tiles that subdivide workload data (tensors), and \emph{dataflow} as the traversal of tiles in space/time~\cite{maestro, timeloop, zigzag, tileflow, set}. While the mapspace we capture using dataplacement has been explored in ZigZag, the dataplacement concept we define in this paper enables a novel analysis of the impact of dataplacement and dataflow combinations on data reuse. This information lets us identify and prune suboptimal dataflows that unnecessarily refetch tiles ($10^{15}\rightarrow1$) and tile shapes that are larger than necessary for reuse ($10^{22}\rightarrow10^5$). Moreover, our representation of dataplacement allows the search to be ordered such that we explore dataplacements before dataflows and tile shapes. This order is more efficient because there are far fewer dataplacements (16) relative to dataflows ($10^{15}$) and tile shapes ($10^{22}$).

\tanner{Assume at this point we've said we've picked a dataplacement and are going to pick dataflows. How can we write the paragraph below such that it doesn't need to define loop nests?}

\tanner{define dominated. key is that we can compare on all attributes (Pareto-better?)}


\tanner{Struggling a bit with the above because the word "reuse" isn't yet defined}

Finally, dataplacement and dataflow determine qualitative information about tiles (``what", ``where", and ``when"), while tile shapes affect only quantitative information (``how big" and ``how many"). This means that most of the modeling complexity is in dataplacement and dataflow. In fact, for a given dataplacement and dataflow, accesses, energy, and latency can be calculated using functions of simple, easy-to-reason-about operators (\eg multiply, add, min, ceiling), which lets us reason about these functions to find tile shape choices that dominate others ($10^6\times$ reduction). After generating these simple functions, we can also evaluate them instead of running a full model to speed up the search process by $> 100\times$.



We implement these ideas in the Turbo-Charged Mapper (TCM), which fully searches the mapspace, making it the first mapper to guarantee finding optimal mappings in feasible runtime (seconds to minutes). We also compare to prior works that trade off optimality for speed~\cite{zigzag,loma,timeloop}, showing that, given similar runtime, this work finds mappings with $1.3-16,000\times$ lower energy-delay-product (EDP).

In summary, we make the following contributions:
\begin{itemize}
    \item We introduce the concept of \dataplan that defines which tiles are stored in each level of the memory hierarchy. \Dataplan clearly shows the relative sizes and lifetimes of each tile in memory.
    \item We show how to use \dataplan to identify and prune suboptimal dataflows and tile shapes.
    \item We introduce \emph{partial-tile-shape-pruning}, which, for a given \dataplan and dataflow, can identify and prune suboptimal tile shapes.
    \item We introduce a fast model that is compiled once for each \dataplan and dataflow, and is then $1000\times$ faster than prior fast models.
    \item We integrate these contributions into \mappername, the first mapper to guarantee finding optimal mappings in feasible runtime. TCM fully searches the mapspace, pruning only suboptimal mappings.
    \item We demonstrate that, compared to prior mappers, TCM improves accelerator energy-delay-product by $1.2-6.5\times$ while simultaneously reducing mapping search time by $1000\times$ (5 hours $\rightarrow$ 17 seconds).
\end{itemize}

%% file: sections/2.background.tex
\section{Background}

\subsection{Einsum Notation}
We describe DNN computations (\eg matrix multiplications, convolutions, activation functions) using the \emph{Einsum notation}. In the Einsum notation, each multi-dimensional data is referred to as a \emph{tensor} and its dimensions (\eg channels in a feature map) as \emph{ranks}. Valid coordinates in that rank (\eg channels 0 through 2) are the \emph{shape} of the rank\footnote{Coordinates in all ranks in this paper start with index 0, so we specify the exclusive upper-bound as the shape. For example, shape 3 denotes $\{0,1,2\}$.}, and the shapes of ranks in a tensor are collectively referred to as the tensor's shape.

Einsums are written like sums, but omit the sum symbol and write index expressions as subscripts. For example, consider the matrix multiplication where \emph{rank variables} $m,k$, and $n$ index into tensors $A$, $B$, and $Z$. Written as an Einsum, we make implicit the summation over $k$ (in general, we sum over all variables not present in the equation's left-hand side):

\begin{equation} \label{eq:mm}
    \begin{split}
        \text{Sum: } Z[m,n] &= \sum_{k}{A[m,k] \times B[k,n]} \\ 
        \text{Einsum: } Z_{m,n} &= A_{m,k} \times B_{k,n}
    \end{split}
\end{equation}

\subsection{Mappings in LoopTree Notation}
A \emph{mapping} schedules, in space and time, a workload's computation steps and all operations in them~\cite{eyeriss,timeloop,efficient_processing_of_dnn}. This paper represents mappings with \emph{LoopTrees}~\cite{looptree,looptree_thesis}, which clearly show mapping choices and their consequences.

Fig.~\ref{fig:placeholder}(a) shows an example mapping as a LoopTree for the Einsum in Eq.~\ref{eq:mm}. We show LoopTrees for an accelerator with two memory levels: DRAM and Global Buffer (GLB). LoopTrees consist of:

\begin{figure}
    \centering
    \includegraphics[width=\linewidth]{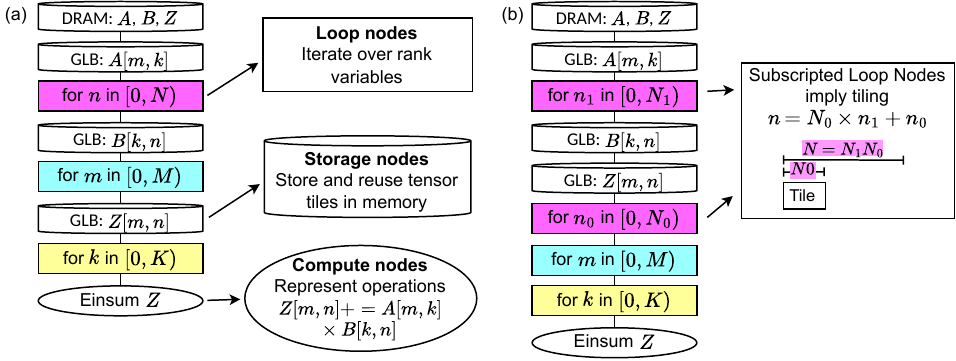}
    \caption{(a) An example LoopTree, the different types of nodes, and their meaning. (b) An example LoopTree with tiling.}
    \label{fig:placeholder}
\end{figure}

\begin{itemize}
    \item \emph{Loop nodes}, rectangles in the LoopTree, represent nested \texttt{for} loops that iterate over rank variables in the workload. Multiple loops over a rank variable imply tiling, and subscripts indicate iteration over tiles and within tiles (\eg in Fig.~\ref{fig:placeholder}(a), the loop over $m1$ iterates over tiles, the loop over $m0$ iterates within a tile, and we index with $m = m_1 M_0 + m0$).

    
    \item \emph{Storage nodes}, cylinders in the LoopTree, represent tensor tile storage. Each tile's size is determined by the loops below the storage node (\eg Fig.~\ref{fig:placeholder}(a) and (b) show that the GLB $A$ storage node must keep a $K_0\times M_0$ tile). The number of tiles fetched (including refetches) is determined by the loops above the storage node (\eg in Fig.~\ref{fig:placeholder}(a), the $A$ tile in GLB is fetched $M_1K_1$ times).
    
    

    \item \emph{Compute nodes}, ovals in the LoopTree, represent the computation that is performed. In this example, they represent the multiply-accumulate operations in the Einsum.
\end{itemize}

To complete the terminology:
\begin{itemize}
    \item \emph{Dataflow} is the relative ordering of operations, which in LoopTree notation is the loop nodes and their order.
    \item \emph{Tile shapes} are the shape of tiles in each rank, which in LoopTree notation are the loop bounds.
    \item \emph{Dataplacement} is the tensor tiles that are held in memory levels, which in LoopTree is the storage nodes and their order. Section~\ref{sec:dataplacement} discusses more details.
\end{itemize}

\subsection{Finding Optimal Mappings is Challenging}
Finding optimal mappings requires a mapping search, as there is no known direct solver. This is because each mapping choice impacts many factors (\eg parallelism, bandwidth utilization, capacity utilization), which affect final energy and latency.

Mapspace search is challenging because the mapspace is very large. Only a tiny fraction of the mapspace can be evaluated within a feasible runtime, even with the fastest models~\cite{timeloop,maestro,zigzag,looptree,tileflow,set}. Then, the challenge is determining which mappings out of the mapspace should be evaluated to find the best mapping possible.

\subsection{The Mapspace is Large and Dataplacement Enables Pruning}

\insightbox{Exploring the mapspace involves exploring the space of dataflows, dataplacements, and tile shapes. There are orders-of-magnitude fewer dataplacements, so we enable the most pruning by exploring dataplacement first.}

\begin{figure}
    \centering
    \includegraphics[width=0.95\linewidth]{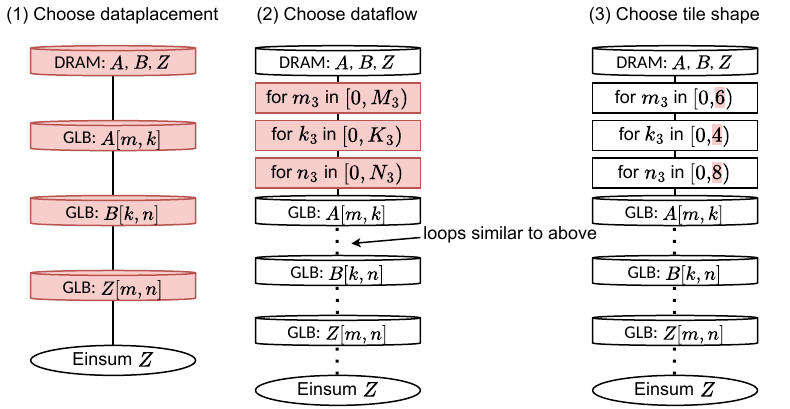}
    \caption{Constructing a mapspace for the Einsum in Eq.~\ref{eq:mm} by choosing dataplacement, dataflow, then tile shape. The part of the mapping chosen in each step is highlighted in red. In (2) and (3), only some choices are shown; similar choices must be made for parts in ellipses.}
    \label{fig:mapspace_construction}
\end{figure}

We can construct the mapspace by making a series of decisions: picking a dataplacement, a dataflow, and then tile shapes. Fig.~\ref{fig:mapspace_construction} illustrates this process. (1) We choose a dataplacement $dp \in DP$, which is the order of storage nodes and whether to include them (\eg whether to keep tensor $I$ in GLB). Storage nodes for different tensors can be reordered freely, but storage nodes must follow memory hierarchy order (\eg DRAM has to be above GLB)\footnote{TCM can optionally relax this assumption, making the memory hierarchy constraint per-tensor (\eg (GLB, input) storage node must go above (register, input) storage node, but may be below (register, output) storage node). This expands the mapspace, and TCM is still optimal in the larger mapspace.}. It is also common to have only one storage node for the highest memory level (\eg DRAM) that stores every tensor in entirety. (2) loops over every rank variable are placed between every pair of adjacent two storage nodes, and we choose a dataflow $df \in DF$, which is the choice of loop orders. (3) we choose tile shapes $ts \in TS$, which are the loop bounds. Following this process, the mapspace size is:
\begin{equation}
    |\text{Mapspace}| = |DP| \cdot |DF| \cdot |TS|
\end{equation}

The number of options varies significantly across different choices, and $|DP|$ tends to be smallest by a wide margin. The reason for this can be seen in Fig.~\ref{fig:mapspace_construction}; there are many more loops than storage nodes, making the spaces of loop orders and loop bounds much larger than the space of storage node orders. For example, mapping the attention matrix computation step in GPT-3 to a TPU-like architecture, $|DP|=16$, $|DF| \approx 10^{15}$, and $|TS| \approx 10^{22}$.

Leveraging this variation, we maximize pruning potential by focusing on reducing $|DF|$ and $|TS|$. The following sections describe how we prune dataflows when given a dataplacement choice and how we prune tile shapes given dataplacement and dataflow choices. For the same example above, $|DF|_{\text{pruned}} =1$ and $|TS|_{\text{pruned}} \approx 10^5$, representing a 32 order of magnitude pruning rate overall.

\subsection{Prior Works Use Heuristics}\label{sec:prior_works}
To manage the large mapspace, prior work has proposed various heuristics~\cite{zigzag,timeloop,loma} and metaheuristics~\cite{tileflow,gamma_mapper,spotlight,set}, but these approaches cannot guarantee finding an optimal mapping. Heuristic approaches use simple-to-evaluate rules (\eg maximize parallelism) to narrow down which mappings to evaluate. However, these rules only capture some of the factors that determine latency and energy, and ignore important trade-offs. For example, increasing parallelism can result in more memory accesses and, when bandwidth is limited, increase latency instead.

Prior mappers also use metaheuristics such as sampling~\cite{timeloop}, genetic algorithms~\cite{tileflow,gamma_mapper}, Bayesian optimization~\cite{spotlight}, simulated annealing~\cite{set}, and Monte Carlo Markov chains~\cite{flexflow}. However, while these approaches yield better-than-random results, they have no guarantee of finding the optimal mapping.

While some prior works include optimality-preserving pruning techniques for dataflows~\cite{zigzag, sunstone}, without prior knowledge of the dataplacement, this pruning is less effective than ours, and these works still need heuristics to search the mapspace.

Other prior works have explored a subset of the dataplacement space. ZigZag~\cite{zigzag} first picks a dataflow, then applies a dataplacement to that dataflow. However, this procedure makes dataplacement and dataflow inextricably linked. We introduce dataplacement as a separable choice, which lets us pick dataplacement first, enabling analysis and highly effective mapspace pruning based on dataplacement.

%% file: sections/3.body.tex
\section{Defining and Interpreting Dataplacement}\label{sec:dataplacement}

\insightbox{Dataplacement describes how tiles are stored in memory, and dataplacement offers new insights, letting us better interpret (and prune) mappings.}

First, we precisely define dataplacement, how we specify it in the LoopTree notation, and its impact on data reuse.

\subsection{Understanding and Specifying Dataplacement}

\emph{Dataplacement}, which determines the tensor tiles present in each memory level, is represented in the LoopTree notation as storage nodes and their ordering. Existence of a storage node implies whether a tile is stored (\eg a storage node ``GLB keep $A$" says the GLB keeps tiles of Tensor $A$), and storage node order affects tile shapes, how long tiles live (\ie their \emph{lifetimes}), and the amount of reuse.

\begin{figure}
    \centering
    \includegraphics[width=\linewidth]{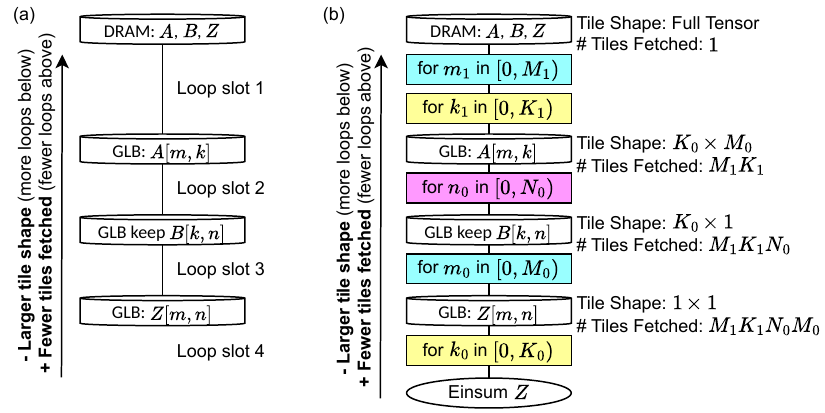}
    \caption{(a) An example dataplacement and slots where loops may be inserted. (b) An example mapping using the dataplacement, including tile shapes and numbers of tiles fetched. Dataplacement shows the trade-off between tile size and number of tiles fetched: Storage nodes higher in the dataplacement reduce fetches, but can increase tile size.}
    \label{fig:mapping_example}
\end{figure}

Fig.~\ref{fig:mapping_example} shows how dataplacement affects tile shapes, lifetime, and reuse. Fig.~\ref{fig:mapping_example}(a) shows the dataplacement alone, while Fig.~\ref{fig:mapping_example}(b) shows a full LoopTree, which includes loops inserted into slots between storage nodes. The dataplacement enforces two invariants on the lifetime and shape of stored tiles:

\emph{Invariant 1: Storage nodes higher in the dataplacement have the same-or-longer lifetime than those lower in the dataplacement}. For example, the ``GLB keep $A$" storage node in Fig.~\ref{fig:mapping_example}(b) is alive for the iterations of all loops in slots 2, 3, and 4; while the ``GLB keep $Z$" storage node below is only alive for loops in slot 4, and new tiles are fetched for each iteration of the loops in slots 2 and 3. This results in fewer fetches, and thus more reuse, for the ``GLB keep $A$" storage node.

\emph{Invariant 2: Storage nodes higher in the dataplacement have the same-or-larger shape than those lower in the dataplacement}. (Note that shape comparisons apply only to ranks shared between the stored tensors.) For example, ``GLB keep $A$" storage node Fig.~\ref{fig:mapping_example}(b) stores a $K_0\times M_0$ tile of Tensor $A$, while ``GLB keep $Z$" stores a $K_0\times1$ tile. This invariant is again because higher storage nodes are above more loops than lower storage nodes, and in the LoopTree notation, the tile shape is the product of loop bounds of relevant loops under the storage node.

Increased lifetime and tile shape mean that higher storage nodes often have more reuse opportunities than lower storage nodes. However, only loops that have reuse opportunities will result in reuse. For example, the tile specified by ``GLB keep $A$" reuses data across iterations of the $n_0$ loop, but not the $k_0$ loop, because each $k_0$ iteration accesses a different value of $A$.

\subsection{Dataplacement Enables Mapping Insights}
Dataplacement offers new insights into how mappings use memory resources to reuse data and reduce refetches. In particular, dataplacement provides a notion of \emph{reuse priority}, which determines the relative amount of memory resources to be used to get more reuse of different tensors.

Traditionally, hardware designers analyze reuse by looking at the dataflow, inferring how tensor tiles are reused based on the order of operations. A dataflow-centric analysis (\eg categorizing dataflows by stationarity~\cite{eyeriss,eyeriss_isca,efficient_processing_of_dnn}) would look at all loops that exist between any two memory levels. For example, in Fig.~\ref{fig:mapping_example}, the DRAM-GLB dataflow is based on the uppermost four loops (between the DRAM storage node and the lowest GLB storage node), and the dataflow is $B$-stationary because the innermost of these loops, the $m_0$ loop, repeatedly accesses the same elements of $B[k,n]$. Therefore, we can store and reuse $B[k,n]$ for multiple iterations.

Analyzing mapping through the lens of dataplacement offers two new insights. First, the dataplacement shows clearly which tiles are kept in the memory longer and the relative shape of the tiles. For example, in Fig.~\ref{fig:mapping_example}, the dataplacement implies that the tile of $A[m,k]$ is kept in GLB for longer and has a larger shape, followed by tiles of $B[k,n]$ and $Z[m,n]$ in order. Second, as a corollary, the dataplacement shows the \emph{relative} priority of tensors. In our example, $A[m,k]$ has the highest reuse priority, followed by $B[k,n]$ and $Z[m,n]$.

Reuse priority provides an intuition for the benefits of the ability to specify a storage node for each tensor (also referred to as \emph{uneven mappings}~\cite{zigzag}), which have been shown to reduce energy and latency~\cite{zigzag,looptree}. Works that focused only on dataflow and implicitly assume a dataplacement (often placing storage nodes for all tensors at the same place)~\cite{timeloop,sunstone,tileflow}, missed this optimization because they only reuse the tensor(s) that are kept stationary for a given dataflow.
For example, in Fig.~\ref{fig:mapping_example}, placing storage nodes for $A[m,k]$ at the same level as $GLB: Z[m,n]$ eliminates all reuse of $A$.

Because it provides both a clear way to interpret mappings and a richer space to explore, dataplacement is important to our understanding of mapping.




\section{Speeding Up Mapping Using Dataplacement}

\insightbox{Using dataplacement, we identify multiple opportunities to prune the mapspace by orders of magnitude. These opportunities come because dataplacement tells us which tiles are stored in each memory level, so we can prune mappings that redundantly re-fetch tiles, store too-large tiles, or lead to identical tile movement as an already-explored mapping.}

We show how dataplacement lets us identify and prune many dataflows, loops, and tile shapes that are guaranteed to be worse-than or equal-to the optimal choices. Then, we show how we can use dataplacement to speed up the model.

\subsection{Redundant Dataflow Pruning}\label{sec:dataflow_pruning}
\insightbox{Many mappings vary in compute order but lead to identical tile movement. These are redundant and can be pruned.}

Knowing dataplacement, we can reduce the number of mappings to evaluate by identifying redundant dataflows. 

Many dataflows are redundant and do not need to be evaluated. In general, dataflows that differ only in the order of loops between the same storage nodes result in exactly the same tile shapes and data movement. For example, in Fig.~\ref{fig:mapping_example}(b), permuting the top two loops would not change tile shapes or the number of fetches. Thus, we do not need to explore orders of loops that are not separated by storage nodes.

This pruning represents significant improvement over prior dataflow explorations~\cite{timeloop,zigzag}, which explore all loop orders instead of exploring only non-redundant dataflows for each dataplacement. There are fewer storage nodes than loops ($\leq3$ storage nodes per memory level, versus $3-7$ loops), leading to far fewer storage node orders than loop orders (less than $10^3$ versus more than $10^7$). Because there are few non-redundant dataflows for each dataplacement, we reduce the number of choices we must explore significantly.

The one exception to this pruning rule is if the rank variable is part of an expression in the index (\eg $p$ in a convolution $Z_{p} = A_{p+r} \times W_{r}$), sometimes referred to as a \emph{partially relevant} rank variable~\cite{zigzag}. In such cases, a loop over $p$ directly under a storage node for $A$ allows only a line buffer to be held in storage for $A$, which reduces memory requirements~\cite{looptree,timeloop,zigzag,eyeriss,fusedcnn}. Such optimization is only applied to the loop directly under the storage node. Thus, when such rank variables exist, we explore loop orders that leads to a unique partially-relevant loop beneath applicable storage nodes (\eg for $Z_{p} = A_{p+r} \times W_{r}$, for each $A_{p,r}$ storage node, there is a choice to put the $p$ loop highest or the $r$ loop highest).



\subsection{Non-Helpful Loop Pruning}\label{sec:loop_pruning}
\insightbox{Many mappings do not help tile movement, but cost increased tile movement or tile size. These are suboptimal and can be pruned.}

Knowing dataplacement, we can identify and remove loops that have no benefit but cost more data movement and/or larger memory requirements.

\begin{figure}
    \centering
    \includegraphics[width=0.82\linewidth]{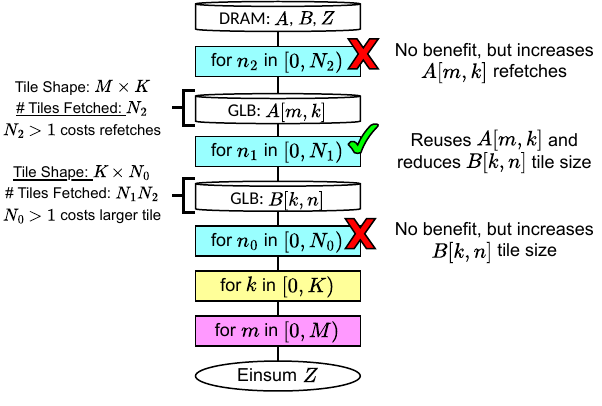}
    \caption{A mapping with non-helpful loops. Notice that the $n_2$ and $n_0$ loops both have costs (more fetches, more tile size) but no benefits (lower fetches, lower tile size). The $n_1$ loop is strictly better than either. Loop pruning will eliminate the $n_2$ and $n_0$ loops while keeping the $n_1$ loop.}
    \label{fig:non_helpful_dataflow}
\end{figure}

For example, in Fig.~\ref{fig:non_helpful_dataflow}, the uppermost $n_2$ loop is not helpful because it increases refetches of $A$ without additional benefit. Each iteration of the $n_2$ loop accesses the \emph{same} tile of $A[m,k]$. Therefore, there is a reuse opportunity for $A[m,k]$. If we remove the loop (equivalent to setting $N_2=1$), we get more $A[m,k]$ reuse and reduce the number of $A[m,k]$ tiles fetched from DRAM without increasing memory requirements.

The lowermost $n_0$ loop is not helpful because it increases $B[k,n]$ tile size (larger memory requirement) without additional benefit. Each iteration of the $n_0$ loop accesses a different element $B[k,n]$, which means that there is no reuse opportunity. Therefore, keeping multiple $n_0$ in GLB results in a larger tile, but does not reduce the amount of data fetched from DRAM. If we remove the $n_0$ loop, we reduce the $B[k,n]$ tile size without increasing data movement.

In summary, we can remove loops that do not index into the tensor in the storage node below or do index into the tensor in the storage node above (see Table~\ref{tab:rank_variables}).

This analysis significantly decreases the number of loops to consider. In matrix multiplications, the number of loops per storage node goes $3\rightarrow1$. For more complex workloads, the reduction is greater (\eg $6\rightarrow2$ per storage node in our GPT-3~\cite{gpt3} experiments). Since the number of tile shapes (ways to factorize workload shape to loop bounds) and dataflows (orders of loops) increases exponentially with the number of loops, this pruning dramatically decreases mapspace size.

We note that ZigZag~\cite{zigzag} includes similar observations on helpful and non-helpful loops, calling it the Loop Relevance Principle (LRP). Our analysis on whether a loop is helpful matches ZigZag's. The difference is that ZigZag uses this analysis to prune dataplacements (create a dataflow (loop nodes) first; only insert storage nodes such that the surrounding loops are helpful), while we use it to prune loops (create a dataplacement (storage nodes) first; only insert helpful loops). Pruning loops is much more effective as it reduces the dataflow and tile shape space, which is much larger than the dataplacement space ($10^{10}-10^{25}$ versus $10-1000$).


\begin{table}
    \centering
    \begin{tabular}{l l l}
        Reuse tensor above   & Reuse tensor below          & Pruned? \\
    \toprule
        Yes                  & No                          & No \\
        No                   & *                           & Yes \\
        *                    & Yes                         & Yes \\
    \toprule
    \end{tabular}
    \caption{Rules to prune a loop based on whether it reuses the tensor in the storage node above/below it. Asterisk (*) means any.}
    \label{tab:rank_variables}
\end{table}


\subsection{Partial Tile Shape Pruning} \label{partial_tile_shape_pruning_1}
\insightbox{For a given dataplacement, the model reduces to simple expressions for the size and number of fetches of each tile. We can analyze these expressions to do additional pruning.}

Hardware performance models, even analytical ones, generally involve complex calculations, but we can simplify the model and conduct trade-off analysis if we fix certain design choices. For example, while it would be difficult to write a single expression that calculates DRAM accesses for arbitrary mappings, the number of DRAM accesses by Tensor $B$ elements in Fig.~\ref{fig:mapping_example} and the GLB usage of $A$ are:
\begin{equation}\label{eq:dram_accesses_B}
\begin{split}
    \text{DRAM Accesses}_B = M_1 \cdot K \cdot N \\
    \text{GLB Usage}_A = M_0 \cdot K_0
\end{split}
\end{equation}

In fact, given dataflow and dataplacement, we can generate expressions for the usage and accesses to every memory level, as well as final metrics such as energy and latency.

Using these expressions, we can identify and prune suboptimal mappings even before knowing all the tile shapes. A partially-chosen set of tile shapes can be pruned if we can show that it will result in worse metrics regardless of future tile shape choices (\eg the tile shape along a dimension may so severely underutilize an array of computation units that it results in high latency regardless of the tile shape along other dimensions). Finding such pruning candidates requires complex analysis, which is described in Section~\ref{make_pruning_criteria}.

\subsection{Speeding Up Modeling via Currying}
\insightbox{For a given dataplacement, the model reduces to simple expressions for the size and number of fetches of each tile. We can avoid long-running model calls by evaluating these expressions instead.}

Explicit dataplacement also enables much faster modeling runtime. We can compute the mathematical expressions as in the previous section first for a combination of dataflow and dataplacement. Then, we use those expressions to explore various tile shapes, which is now much faster because it only involves numerical substitutions and arithmetic operations. In other words, we separate the model such that the input can be given one at a time (\ie we \emph{curry} the model). Moreover, each mathematical expression can be used for the entire tile shape exploration. Thus, we trade off the slightly more expensive symbolic analysis of the impact of dataplacement and dataflow to obtain a faster model that we can reuse for all tile shapes.

%% file: sections/4.body2.tex
\section{The Turbo-Charged Mapper (TCM)}
\insightbox{TCM iteratively follows an \emph{enumerate-prune} strategy, repeatedly pruning as it generates dataplacement, then dataflow, then tile shape choices. This lets TCM prune the search space by many orders of magnitude, making it tractible to fully search the remainder and find the optimal mapping.}

\begin{figure}
    \centering
    \includegraphics[width=\linewidth]{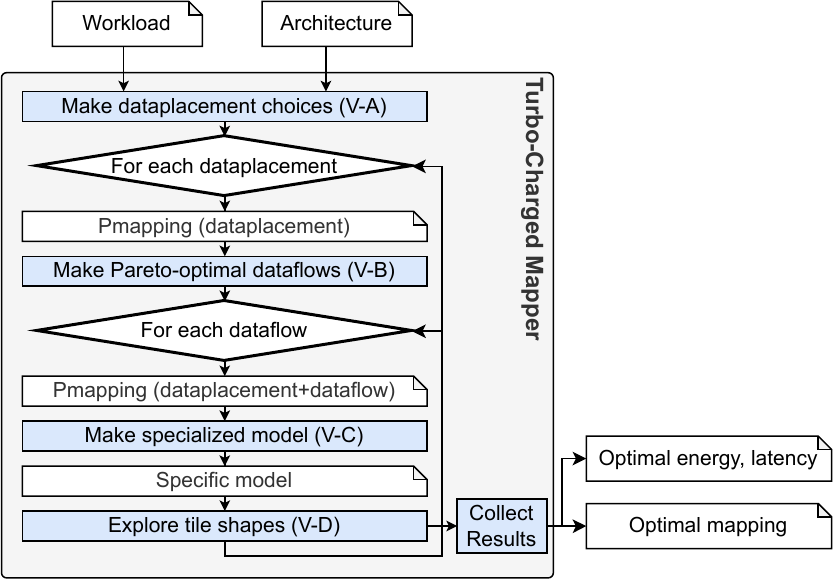}
    \caption{Overview of \mappername. Each mapper process (in blue) is explained in detail in the sections in parentheses.}
    \label{fig:overview}
\end{figure}

In this section, we describe the Turbo-Charged Mapper (TCM), which implements the ideas in Section~\ref{sec:dataplacement}. TCM, shown in Fig.~\ref{fig:overview} is the first to output optimal mappings in a feasible runtime.

Optimal mappings are guaranteed because TCM only prunes choices that are known to be suboptimal. Meanwhile, when exploring at a given step, it explores every possible choice that may be optimal (\ie it does not miss anything).

TCM includes four key steps:

\begin{itemize}
    \item \textbf{Make dataplacement choices}. This is equivalent to making every possible ordering of storage nodes in the LoopTree. This step outputs a partial mapping (\emph{pmapping}), which is a mapping with only some attributes specified.
    \item \textbf{Make dataflow choices for each dataplacement choice}. This step reduces mapspace size by only generating Pareto-optimal dataflows.
    \item \textbf{Make a tile-shape-only model}, which fixes dataflow and dataplacement to take only tile shapes as input.
    \item \textbf{Explore tile shapes}. This uses the tile-shape-only model. As it chooses tile shapes, it prunes known-suboptimal choices to keep the search size tractable.
\end{itemize}

The four steps are detailed in the following subsections.

\subsection{Making Dataplacement Choices}
\mappername explores all dataplacement choices, which involves making and ordering storage nodes. Making storage nodes means that we decide, for each memory level and tensor, whether to keep a tile of the tensor in the memory level (\ie whether the storage node exists). \mappername generates all possible combinations of storage nodes (except those that are not supported by the hardware). Then, for each combination, \mappername generates all possible hardware-supported storage node orders. Storage order must also obey the hardware memory hierarchy, meaning that we can not have the $L1: A[m,k]$ storage node higher than the $L2: A[m,k]$ storage node, else the mapping would use $L1$ as a backing store for $L2$.

Users may constrain the dataplacement as well. This may be useful if a memory is used to store a particular tensor (\eg WeightBuffer can only hold weights).

\subsection{Making Pareto-Optimal Dataflows}
For each dataplacement choice, \mappername explores all non-redundant dataflow choices. We first explore ordering of the loops in each slot, then all combinations of per-slot choices.

Between any two storage nodes, we determine the set of loops over rank variables that would be beneficial to insert between any two storage nodes using the principles discussed in Section~\ref{sec:loop_pruning}. We then partition the rank variable (\eg $n\rightarrow n_0n_1$) and insert a loop between the two storage nodes. In cases where there is no lower storage node (\ie below the last storage node, above the compute node), we omit the reuse-tensor-below check. In cases where the upper storage node is the outermost backing storage (\eg DRAM) we omit the reuse-tensor-above check since loops may not be placed above this backing storage. Given the set of loops between storage nodes, we explore choices of which loop is directly under the storage node above, and we choose a canonical order (\eg alphabetical) for the rest of the loops because the order does not matter (see Section~\ref{sec:dataflow_pruning}).

\subsection{Making the Tile-Shape-Only Model}

Like prior work~\cite{timeloop, zigzag, loma}, our model takes in workload, architecture, and mapping specifications as inputs to calculate overall energy and latency. However, our model is much faster for mapping search because it performs the most time-intensive modeling work (analyzing a particular dataplacement and dataflow) a handful of times (tens to hundreds), sharing this effort across all mappings that have the same dataplacement and dataflow (thousands to millions).

To achieve this speed, our model is \emph{curried}, meaning we first give the model dataplacement and dataflow, and it outputs a \emph{tile-shape-only model} that only needs tile shapes as input:

\begin{align*}
\text{\textbf{TileShapeOnlyModel}} &= \text{\textbf{FullModel}}(\text{dataplacement}\!,\text{dataflow}) \\
\textbf{Results} &= \text{\textbf{TileShapeOnlyModel}}(\text{tile shapes})
\end{align*}

The tile-shape-only model is simpler and faster than a full model, as it omits the complexity of dataplacement and dataflow analysis. In fact, the tile-shape-only model can calculate energy, latency, or usage of any on-chip resource using simple sum-of-products and max expressions. Given a dataplacement and dataflow, we calculate the tile size and number of accesses for Storage $s$ with the following:
\begin{multline}
    \text{\textbf{TileSize}}(s)=\prod\nolimits_{l\in\text{Indexing Loops Below}} l.\text{bound} \\
    \text{\textbf{TilesFetched}}(s) = \prod\nolimits_{l\in\text{Loops Above}}l.bound \\
    \text{\textbf{AccessesToAbove}}(s) = \text{TileSize}(s)\cdot\text{TilesFetched}(s)
\end{multline}

For each memory $m$ with storage nodes $\text{Storage}(m)$, we aggregate accesses and tile size:
\begin{multline}
    \text{\textbf{Usage}}(m) = \sum\nolimits_{s\in \text{Storage}(m)}\text{TileSize}(s) \\
    \text{\textbf{Energy}}(m) = \sum\nolimits_{s\in \text{Storage}(m)}\text{Accesses}(m)\cdot\text{AccessEnergy}(m) \\
    \text{\textbf{Latency}}(m) = \text{Accesses}(m)/\text{Bandwidth}(m)  \\
\end{multline}

Finally, we calculate compute latency and energy, then aggregate across components:
\begin{multline}
    \text{\textbf{Latency}}_{\text{Compute}} = \text{Computes}/\text{UtilizedComputeUnits}  \\
    \text{\textbf{Energy}}_{\text{Compute}} = \text{Computes}\cdot\text{PerComputeEnergy}  \\
    \text{\textbf{Latency}} = \max\left( \max\nolimits_{m \in \text{memories}}\text{Latency}(m), \text{Latency}_{\text{Compute}}\right) \\
    \text{\textbf{Energy}} = \text{ComputeEnergy}+\sum\nolimits_{m \in \text{memories}}\text{Energy}(m)
\end{multline}

This model is like other analytical models~\cite{timeloop,cosa}. The expressions here are computed symbolically, forming the tile-shape-only model, which accepts numerical tile shapes to generate numerical latency and energy values. Then, complex calculations (such as finding all loops above a given storage node) resolve to simple variable substitutions.

Finally, the model we discussed is simplified for clarity; \mappername's model supports many other optimizations, including those not supported in prior work~\cite{timeloop,zigzag}. These optimizations can also be modeled as sum-of-product and max expressions\footnote{Supported optimizations include separated and/or combined read/write bandwidths, varied read and write costs for each tensor, varied quantization for each tensor, additional components such as networks and processing stages (\eg quantization), heterogeneous compute units (\eg choose between a MAC array and a scalar engine), and more complex network communication topologies (\eg multicast and reduction networks).}.



\subsection{Exploring Tile Shapes} \label{make_pruning_criteria}
After choosing dataplacement and dataflow, we must explore tile shapes. Concretely, we must choose numerical values for $\text{TileShapes} = \{b_1,...,b_n\}$, where each $b_i$ is a loop bound in the mapping for which we must provide a numerical value. Because each $b_i$ has many choices, and the full space scales exponentially with the number of loops, the full space is very large (up to more than $10^{10}$ choices).

To address this exponential growth, we explore loop bound values for one loop at a time, starting from the innermost loops,\footnote{At each iteration, we explore options for the loop that will remove the most unknowns from objective formulas. This leads to greater pruning because fewer unknowns means that we can perform fewer partitions of objective functions and Pareto prune with fewer criteria.} and prune our tile shape choices as we make them (\ie we want to prune \emph{partial tile shapes}). With full tile shapes, it is straightforward to detect whether one set of tile shapes is better than another: we can simply call the model, $\text{TileShapeOnlyModel}(b_1,...,b_n)$, compare the results, and prune the worse choice. However, partial tile shapes cannot be evaluated with the model, so we create a different \emph{pruning criteria} for partial tile shapes. 

\subsubsection{Pruning Criteria for Partial Tile Shapes}
We generate the pruning criteria from the optimization objectives (\eg latency or energy), but each criterion can be evaluated using information within the partial tile shapes. Moreover, each criterion is marked as ``minimize," ``maximize," or ``cannot compare." A choice of partial shapes is better than another choice if it is better in all ``maximize" and ``minimize" criteria (lower in all ``minimize" criterion and higher in all ``maximize" criterion) and exactly matches in all ``cannot compare" criteria.

To describe how we generate our pruning criteria, suppose we are in the middle of the exploration, and we have generated all loop bounds for loops $\{k_1,...,k_m\}$, but not for loops $\{u_1,...,u_p\}$ (the \underline{k}nown and \underline{u}nknown loop bounds respectively). Before choosing another loop to explore, we prune the choices we have generated (\ie the values for loops $k_i$). Since we cannot fully evaluate the model without numerical values for $u_i$, we create various criteria $C_i$, which are dependent only on known tile shapes, $C_i = C_i(k_1,...,k_m)$.


We create criteria by applying rewrite rules to the TileShapeOnlyModel as starting criteria (one criterion for latency, one for energy, and one for usage of every resource). The most common rule is \emph{partitioning} a criterion. For example, suppose we want to minimize GLB usage by tiles of tensors $A$ and $B$.
\begin{align}
    \text{GLB Usage} &= \text{TileSize}(A) + \text{TileSize}(B) = k_0 + u_0u_1
\end{align}
The value of $k_0$ is known, while $u_0,u_1$ are unknown and independent of $k_0$ (we know the tile size of $A$ but not $B$). Then GLB usage is minimized if we minimize both $A$ tile size ($k_0$) and $B$ tile size ($u_0u_1$). So, we partition the GLB usage expression into two criteria, one for each term, to be minimized.

However, not all criteria need to be considered, and we can apply another rule: \emph{dropping symbols}. In our example, the criterion $u_0u_1$ is independent of $k_0$ and only involves symbols with unknown values. At this point in the exploration, we can ignore $u_0u_1$ because we are only pruning choices for $k_0$. In general, we partition formulas until they are small enough to either evaluate or drop. We can partition minimizations, maximizations, sums, and products.

Sometimes, there are tile shape interactions that must be included in the criteria we evaluate. For example, if we have enumerated tile shape choices for loop $for\ x_i\ in\ [0, X_i)$ but not the next-outermost loop of the same rank variable $for\ x_{i+1}\ in\ [0, X_{i+1})$, then choices for $X_i$ constrain future choices of $X_{i+1}$ because $X_iX_{i+1}$ must evenly divide $X$. In such cases, we include $X_i$ as a ``cannot compare" criterion such that each $X_i$ choice yields a different set of $X_{i+1}$ choices.

Finally, while not listed here for conciseness, there are simple rewrite rules for simplifying expressions such as algebraic simplifications (\eg $\frac{k_1k_2}{k_1} = k_2$) and dropping constants (\eg minimizing $5x$ is equivalent to minimizing $x$).

\subsection{Why TCM is Optimal}
TCM finds optimal mappings because it \textbf{(1)} explores every non-pruned mapping and \textbf{(2)} ensures all pruning is optimality-preserving.

To ensure pruning is optimality-preserving, we look at each pruning strategy employed:
\begin{itemize}
    \item \textbf{Loop Pruning:} We eliminate a loop, and thus any mappings including that loop, because we know that placing the loop somewhere else would yield the same or better memory usage and accesses to every memory level. Therefore, we know that there exists a better mapping in our mapspace (that same mapping with the loop placed elsewhere).
    \item \textbf{Dataflow Pruning:} We avoid exploring a dataflow if and only if another explored dataflow is identical in every possible mapping statistic.
    \item \textbf{Partial Tile Shape Pruning:} We prune a set of tile shape choices if and only if another set of tile shape choices is Pareto-better in all of memory accesses, memory usage, latency, and energy, and would yield the same options for future choices.
\end{itemize}

Since no pruning strategy eliminates a potentially-optimal mapping, TCM keeps all Pareto-optimal mappings. While we cannot predict optimizations that will be invented in the future, our optimality guarantee applies in a superset of the mapspaces of many prior works~\cite{timeloop,zigzag,loma,sunstone,cosa,maestro,orojenesis}.



%% file: sections/5.results.tex
\section{Evaluation}
\insightbox{TCM prunes the search space by many orders of magnitude to quickly find optimal mappings. Prior works are unable to find optimal mappings even given orders-of-magnitude more runtime.}

This section shows that \mappername finds the optimal mapping within a feasible runtime. Because \mappername always includes choices that can lead to optimal mappings in the searched space, the best mapping found by \mappername is optimal. Our first result shows that the pruning and model currying we propose speeds up the exploration enough that \mappername finishes within a feasible runtime (\eg approximately 17 seconds for GPT-3 6.7B). Then, we compare \mappername to prior works, showing that the mapping found by \mappername has significantly lower EDP (geomean 2.1$\times$) given equal runtime budget, and still meaningfully lower EDP (geomean 1.2$\times$) even when prior mappers are allowed 1000$\times$ the runtime.

\subsection{Experiment Setup}
\mappername supports user-configurable hardware and workload configurations. We use two workloads and two hardware configurations. 

\subsubsection{Workloads}
Workloads used are GPT-3 6.7B~\cite{gpt3} and MobileNetV3~\cite{mobilenetv3}. GPT-3 workload includes one decoder layer, including embedding, multi-headed self-attention, and feedforward layers. We label the Einsums are as follows.
\begin{itemize}
    \item Q, K, V: the query, key, and value projection in GPT-3.
    \item Z: the output projection in GPT-3.
    \item QK, AV: multi-head attention computation in GPT-3.
    \item FFA, FFB: feedforward layer in GPT-3.
    \item P: pointwise convolution in MobileNetv3.
    \item D: depthwise convolution in MobileNetv3.
\end{itemize}

\subsubsection{Architecture}
GPT-3 workloads are run on a TPU-V4i-like~\cite{tpu,ten_lessons_tpu} datacenter accelerator, and MobileNetV3 workloads are run on an NVDLA-like~\cite{nvdla,nvidia_blackwell} edge accelerator. TPU-V4i configuration is taken from the papers, and includes compute energy, access energy of each memory level, and bandwidths of each memory level (including DRAM). It includes a 128MB global buffer and four PEs. Each PE has a 4MB local buffer and a $128\times 128$ array of MACs and registers (holding one weight value each). The array multicasts inputs on one dimension and reduces outputs on the other.

The NVDLA-like configuration is taken from public data~\cite{nvdla}. Access energy and bandwidths are modeled using HWComponents~\cite{cimloop,accelergy,cacti}. It includes a 64kb global buffer and a $32\times192$ MAC array that reuses inputs along the 32-long dimension and reduces outputs along the 192-long dimension.

\subsection{Validated Model}
The hardware model is based on the validated LoopTree~\cite{looptree} model and component energies are modeled with HWComponents~\cite{hwcomponents, cimloop, accelergy}. We validate our model against LoopTree, Timeloop, and CiMLoop~\cite{looptree,timeloop,cimloop} and it yields identical results. 

Additionally, our open-source release includes additional models, including five fabricated~\cite{jia,sinangil,wan,wan_ii,wang,wang_ii,colonnade} and four simulated~\cite{albireo,raella,lightning,isaac} chips. Each of these chips is verified under a range of workloads and operating conditions (\eg varied supply voltage) and match energy, area, and latency within 10\% of published data. Information on these models can be found in~\cite{cimloop}.

\subsection{TCM Prunes by Orders-Of-Magnitude} \label{section:pruning_rate}
Here, we show how each of our optimizations affects the number of mappings that must be evaluated. We show results for GPT-3 6.7B prefill with batch 64 and sequence length 65,536. MobileNetV3 is shown with batch 64.

Table~\ref{tab:mapspace_size} shows the total number of mappings, the number of unpruned mappings that must be evaluated by \mappername, and the reduction. Across Einsums, the mapspace size varies ($10^{23}$--$10^{37}$) because the number of ranks and the shape of tensors vary. But even for the largest mapspace, \mappername produces a tractable number of non-pruned mappings to evaluate (at most $10^7$ mappings). In fact, across all Einsums, the pruning rate varies between $10^{19}$--$10^{30}$, generally increasing as the size of the mapspace increases.

\begin{table}
\centering
\begin{tabular}{llrrr} 
& & \multicolumn{3}{c}{$\#$ Mappings (orders of magnitude)} \\
\cmidrule{3-5}
Workload             & Einsum        & Total & Non-Pruned & Reduction \\
\toprule
\textbf{GPT-3}       & V, K, Q, Z    & 36 & 7 & 29 \\
                     & QK, AV        & 37 & 6 & 30 \\
                     & FFA, FFB      & 30 & 7 & 22 \\
\addlinespace[1ex]
\textbf{MobileNetV3} & D0     & 28 & 5 & 24 \\
                     & D1, D2 & 30 & 5 & 25 \\
                     & P0     & 21 & 4 & 17 \\
                     & P1, P2 & 23 & 4 & 19 \\
\toprule
\end{tabular}

\caption{Number of total and non-pruned mappings that must be evaluated by \mappername. Einsums with the same results are on the same row. (\eg Einsums $V$, $K$, $Q$, and $Z$ each have $10^{36}$ mappings before pruning and $10^7$ mappings after). Pruning rate (reduction) increases with larger mapspaces, making the number of non-pruned mappings always tractable.}
\label{tab:mapspace_size}
\end{table}

\begin{figure}
    \centering
    \includegraphics[width=\linewidth]{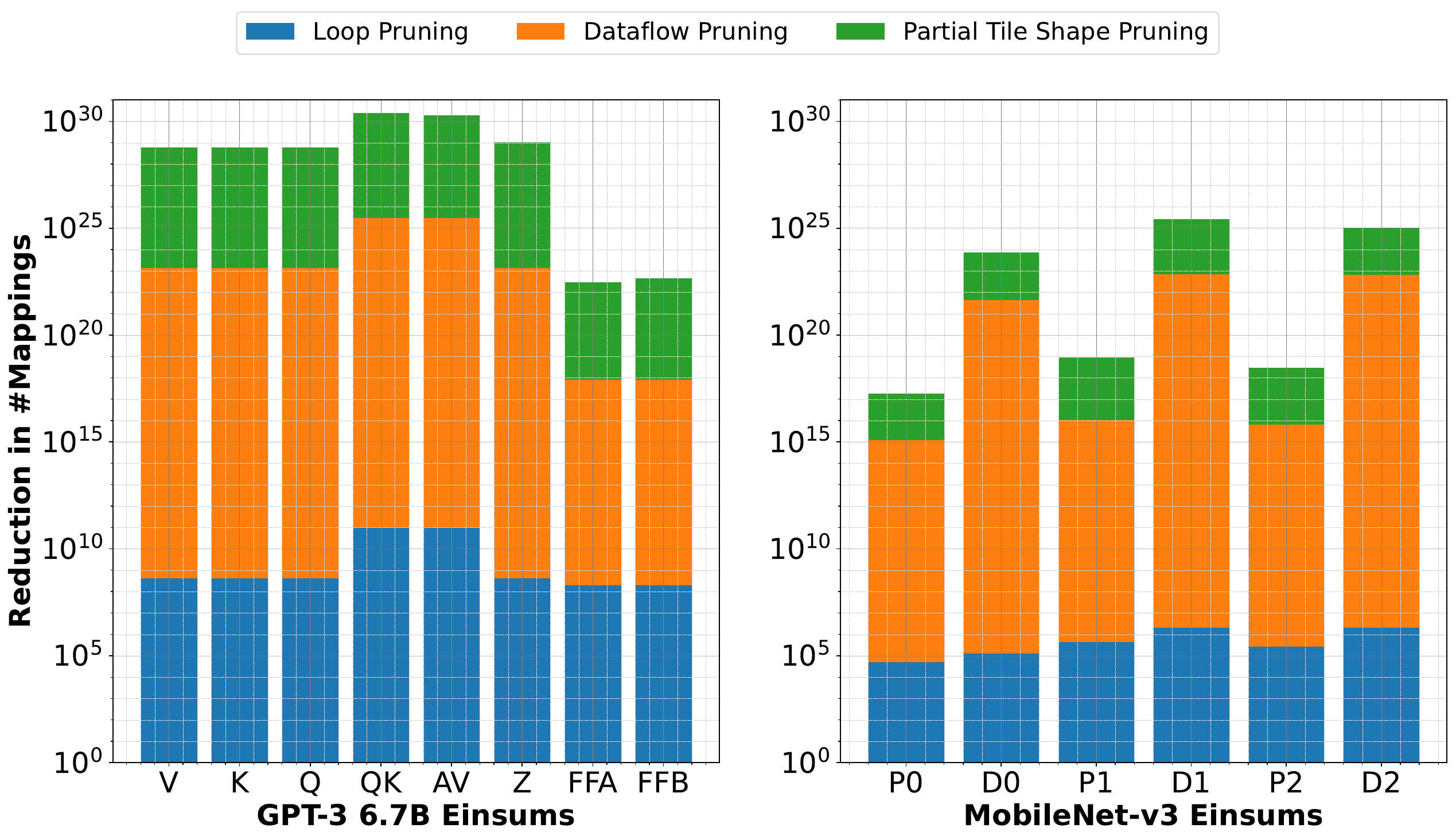}
    \caption{Search size reduction by each of our optimizations. The Y axis is multiplicative and cumulative. For example, the first bar shows that for Einsum $V$, Tile Shape Pruning reduces search size by $\sim 9$ orders of magnitude, Tile Shape and Dataflow Pruning together reduce search size by $\sim 28$ orders of magnitude, and all three together reduce search size by $\sim 29$ orders of magnitude. All optimizations are essential to make this exploration feasible as each reduces search size by orders of magnitude.}
    \label{fig:mapspace_size_reduction}
\end{figure}

To show how \mappername achieves these pruning rates, Fig.~\ref{fig:mapspace_size_reduction} breaks down mapspace size reductions into Tile Shape Pruning, Dataflow Pruning, and Partial Tile Shape Pruning. We can see that Dataflow Pruning is most effective, followed by Tile Shape Pruning, and finally Partial Tile Shape Pruning. All three are essential to make mapspace exploration feasible; the weakest, Partial-Tile-Shape pruning, still reduces search size by up to seven orders of magnitude.

\subsection{Pruning Rate Increases in Larger Mapspaces}
\begin{figure}[b]
    \centering
    \includegraphics[width=\linewidth]{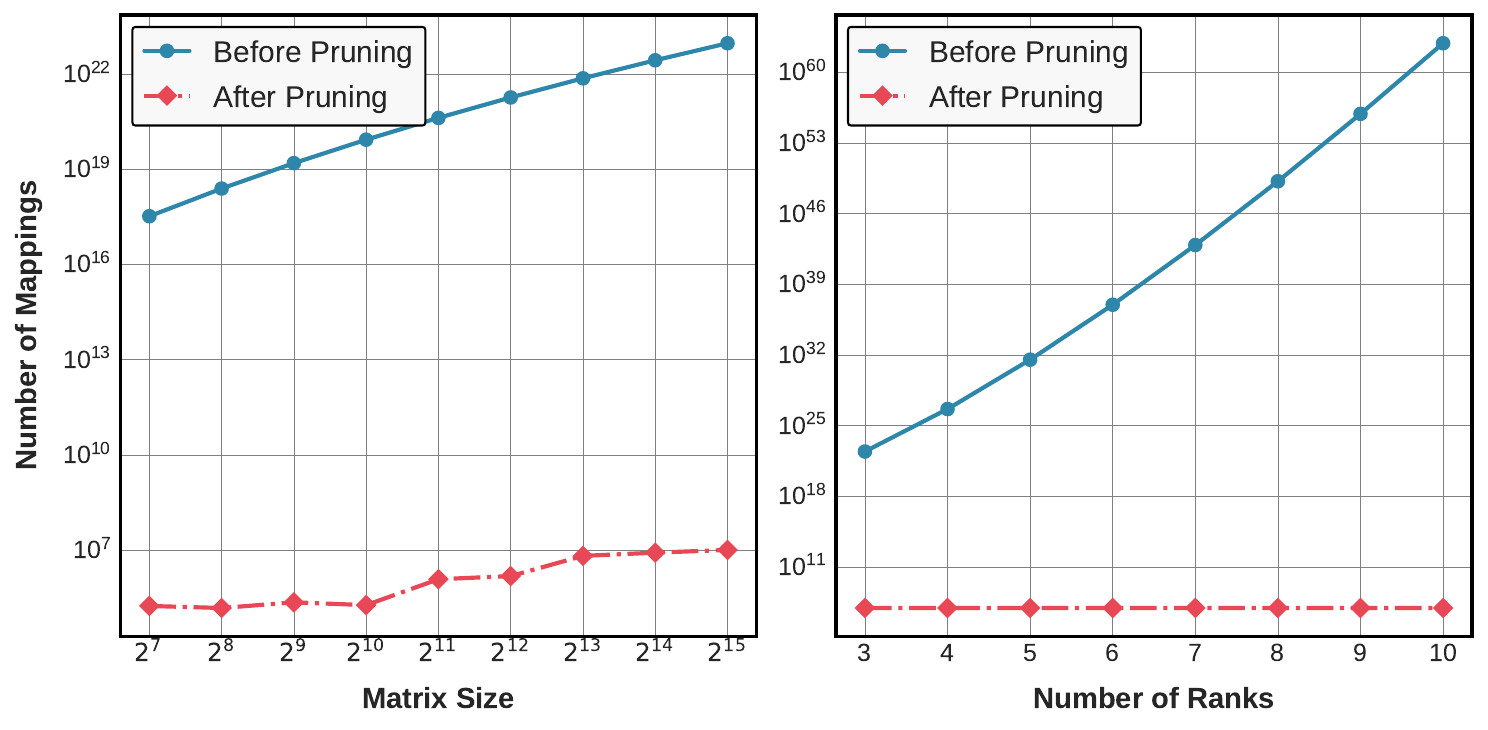}
    \caption{Mapspace size scaling, with and without pruning, for matrix multiplies on a TPU-v4i-like accelerator. (Left) Mapspace size scaling with matrix size, with three total ranks ($M$, $N$, and $K$). (Right) Mapspace size scaling with the number of ranks, with matrix dimensions $M=N=K=16384$ and additional size-1 ranks added to the weight tensor. As mapspace expands, pruning becomes more effective, keeping search size tractable even for very large mapspaces.}
    \label{fig:pruning_rate_scaling}
\end{figure}

As mapspace size increases, pruning rate increases as well, making TCM effective for exploring even extremely large mapspaces. To show why, Fig.~\ref{fig:pruning_rate_scaling} shows how the mapspace size and pruned mapspace size scale for the TPU-v4i-like accelerator running matrix multiplication workload with varied-size matrices and with additional ranks.

As matrix size increases (see Fig.~\ref{fig:pruning_rate_scaling}, left), the number of tile shapes increases because there are more possible tile shapes. However, after pruning, the number of mappings increases at a much slower rate. This is for two reasons. First, loop pruning, in removing loops, eliminates more tile shape options because it removes the possible bounds for those loops. Then, partial tile shape pruning eliminates more tile shapes early on, because more tile shapes tend to be invalid or suboptimal. The one decrease in the plot-- when moving from a matrix size of $2^{11}$ to $2^{12}$-- occurs when the matrices become too large to fit in the TPU-v4i local buffers, and partial tile shape pruning quickly prunes these invalid tile shapes early on.

As the number of ranks increases (right half of Fig.~\ref{fig:pruning_rate_scaling}), the number of mappings increases as the number of loop orders increases. Each additional loop multiplies the number of dataflows (as the number of loop orders increases with the factorial of the number of elements in a set). However, this growth is entirely prevented by Dataflow Pruning, because we explore dataplacements, not loop orders.


\subsection{TCM's Fast Model Yields $400\times$ Speedup}
Here, we show how the curried model affects overall runtime. We test GPT-3 Einsum $QK$, showing the runtime of \mappername while using three different models:


\begin{itemize}
	\item \textbf{Full (Python)}: Our model without currying. This model is a basic, non-optimized Python model.
	\item \textbf{Full (C++)}: An optimized C++ Timeloop~\cite{timeloop} model.
	\item \textbf{Curried}: Our two-step curried model, written in Python. It is first partially evaluated only with dataflow and dataplacement (16 times), then is partially evaluated (2M times) with tile shapes.
\end{itemize}

\begin{figure}
    \centering
    \includegraphics[width=\linewidth]{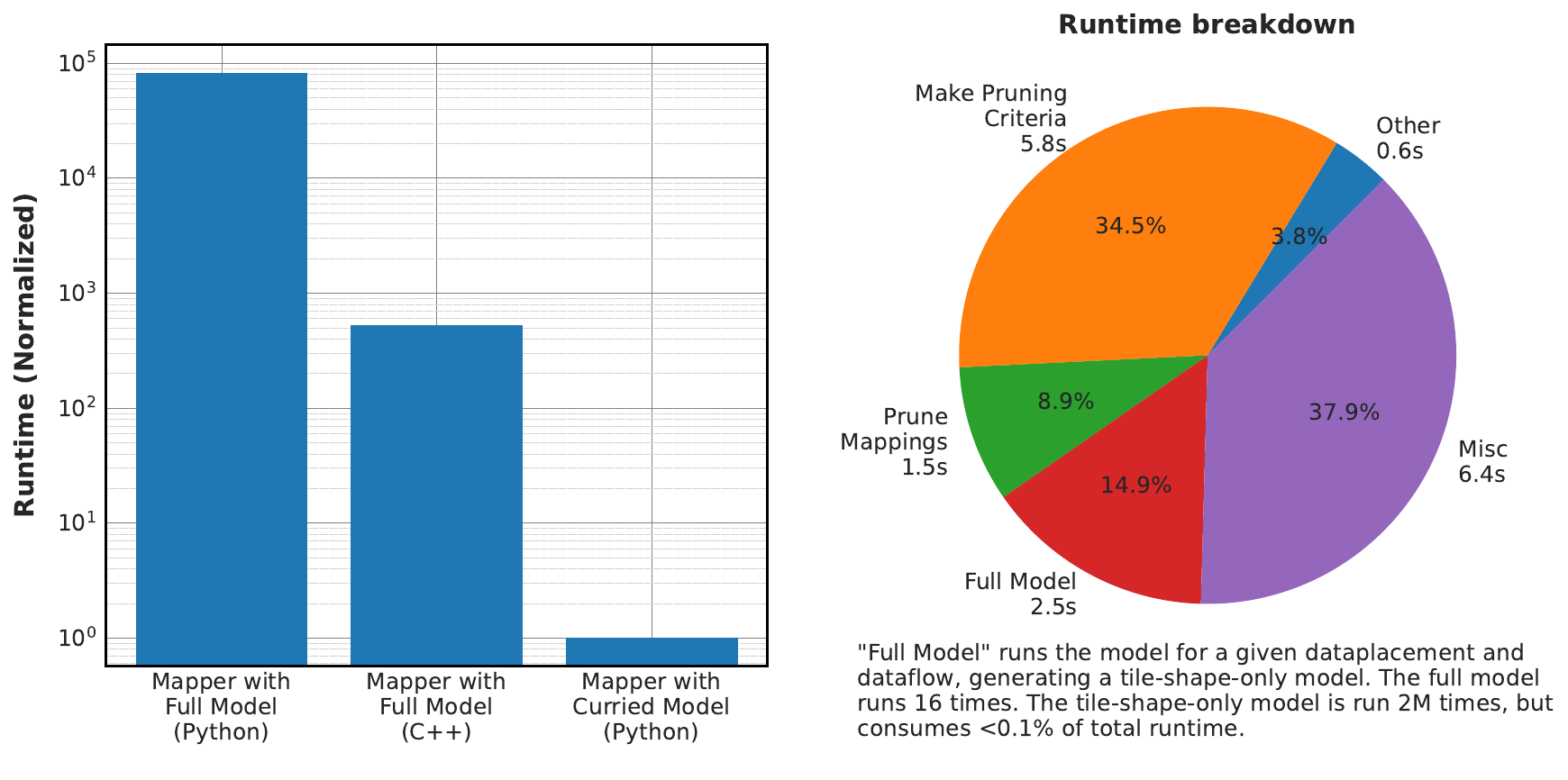}
    \caption{(Left) Model speed comparison. The curried model reduces runtime by $40\times$ and $400\times$ versus full Python and C++ implementations, respectively. (Right) Runtime breakdown with the curried model. The curried model is very fast, and the overall runtime is dominated by other parts of the mapper.
    }
    \label{fig:model_speed_compare}
\end{figure}

Fig.~\ref{fig:model_speed_compare} shows the runtime of the full \mappername while using each model for all evaluations. We see that the curried model is significantly faster than the C++ and Python models, reducing runtime by $40\times$ and $400\times$, respectively.

Fig.~\ref{fig:model_speed_compare} shows the runtime breakdown; with currying, the model is fast enough that runtime is suboptimal by making pruning criteria (Section~\ref{make_pruning_criteria}) and pruning mappings. Partial evaluation of the model for each dataplacement and dataflow, occurring 16 times in this experiment, consumes $<5\%$ of the runtime. Finally, the tile-shape-only model, which has been partially evaluated (going from thousands of lines of Python to a handful of sum-of-product and max expressions), consumed $<0.1\%$ of runtime, despite being called two million times.

\subsection{TCM Is Optimal in Seconds; Prior Works Fail to Find Optimal in Hours}\label{sec:prior_work_comparison}
In this subsection, we compare against prior mappers. We run GPT-3 6.7B Einsum $QK$ on TPU-V4i-like with batch size 64 and sequence length 65,536. We compare to the following mappers:
\begin{itemize}
    \item \textbf{Timeloop}~\cite{timeloop} is a commonly used prior mapper that uses random sampling to explore the mapspace.
    \item \textbf{Timeloop+Hint} reflects a common use case of Timeloop, in which users define mapspace constraints to speed up mapping. We implement a common constraint: the mapping must spatially utilize the full MAC arrays.
    \item \textbf{LOMA}, integrated into the ZigZag~\cite{zigzag} framework, first creates tile shapes, then explores dataflows and dataplacements for these shapes. LOMA constrains mappings to fully utilize PEs and MACs (like Timeloop+Hint).
\end{itemize}

We note that the full-utilization constraints in Timeloop+Hint and LOMA are not guaranteed to be optimal (\eg would yield only invalid mappings if the MAC array is very large and needs tiles too large for the local buffer). However, in this particular case, these constraints preserve the known optimal mapping (found by TCM). 

We first run TCM to completion ($\sim 17$ seconds). We run Timeloop given various timeouts and report the best mapping found. LOMA results are generated by varying LOMA's LPF setting (described in~\cite{loma}), which constrains tile shape choices to reduce runtime. We report the best mapping found by LOMA in the most comprehensive mapspace (highest LPF setting) that finishes within each time limit. Because LOMA is single-threaded, for fairness, we also only use one thread for TCM and Timeloop (both of which can parallelize).

Table~\ref{tab:baseline_runtime_edp_comparison} compared the best EDP found. Within TCM's runtime ($t_0$), all baselines are far from optimal. Timeloop, which samples randomly, has a very high EDP because of severe underutilization of the MAC array (tile shapes that perfectly utilize the array shape are rare). Timeloop+Hint has significantly better EDP, but still picks poor dataflows and tile shapes, often increasing energy and/or latency with excessive off-chip fetches. LOMA quickly achieves a fair EDP ($1.28\times$ optimal) because it includes additional heuristics (always fully utilize spatial units, optimize accesses for one level at a time); however, it improves only marginally with more runtime.

We note that TCM also guarantees optimality, while Timeloop and LOMA can only do so with full exploration of the mapspace (which would require an infeasible $>10^{20}\times$ additional runtime).

\begin{table}
\centering
\begin{tabular}{lcccc}
                       & \multicolumn{4}{c}{Best EDP (normalized) found } \\
    \cmidrule{2-5}
                       Runtime & $t_0$ & $10\times t_0$ & $100 \times t_0$ & $1000 \times t_0$ \\
\toprule
    Timeloop           & 16444 & 1126 & 116 & 6.48 \\
    Timeloop+Hint    &  3.81 & 1.41 & 1.41 & 1.12 \\
    LOMA               & 1.28 & 1.28 & 1.28 & 1.21 \\
    TCM      & 1 & \multicolumn{3}{c}{Completed} \\
\toprule
\end{tabular}
\caption{EDP comparison of TCM and prior mappers given various runtimes (lower is better). TCM finds an optimal mapping in $t_0 = 17$ seconds. Prior mappers fail to find an optimal mapping even when given $1000\times$ the runtime (5 hours).}
\label{tab:baseline_runtime_edp_comparison}
\end{table}

\subsection{TCM Mappings And Hardware Tradeoffs}

\begin{figure}
    \centering
    \includegraphics[width=\linewidth]{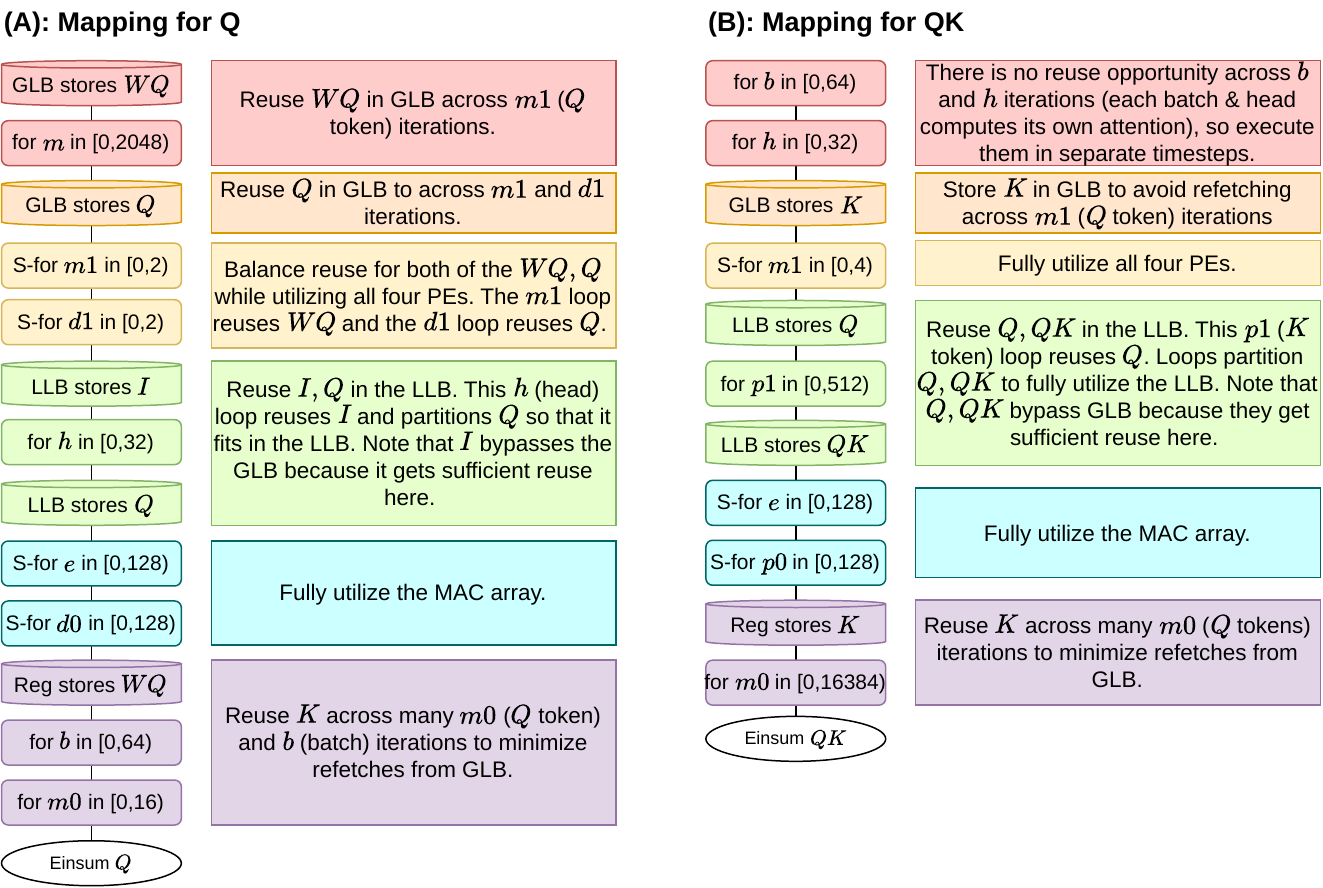}
    \caption{(Left) Mapping for Einsum $Q$, and (Right) Mapping for Einsum $QK$, annotated with the goals the mapper is attempting to accomplish at each level. Memory levels are DRAM (omitted), global buffer (GLB), local buffer (LLB), and reg. S-for denotes spatial loops. Notice that there are many tradeoffs to consider, including capturing reuse to minimize refetches, fully utilizing spatial units, and ensuring that no memories are overprovisioned. 
    }
    \label{fig:mapping_analysis}
\end{figure}

To show the tradeoffs to consider in creating an optimal mapping and why TCM can find them quickly, Fig.~\ref{fig:mapping_analysis} shows the optimal mapping found for GPT3-6.7B~\cite{gpt3} running on TPU-V4i-like with batch size 64 and sequence length 65,536. We show mappings for Einsums $Q$ and $QK$, annotating the goals of the mapper at each level.

Looking at these mappings, we can make multiple observations:
\begin{itemize}
    \item There are many tradeoffs to consider. TCM simultaneously balances the reuse of every tensor at every memory level, often doing things like bypassing memory levels to further reduce memory usage and the memory access energy.
    \item Tradeoffs change across different Einsums. Notice that the mapping for Einsum $Q$ keeps two tensors in GLB and balances reuse of both tensors using spatial loops, while the mapping for Einsum $QK$ keeps only one tensor in GLB. Meanwhile, notice that the mapping for $Q$ puts the batch $b$ loop all the way at the bottom to leverage $WQ$ reuse at multiple levels, while the mapping for $QK$ puts the batch loop all the way at the top because this Einsum has no reuse opportunity across $b$ iterations. 
    \item Tradeoffs interact with one another across multiple memory levels. Notice the $m0$ loop on the bottom of the mapping for $Q$. We'd get maximum $WQ$ reuse in reg if this loop bound was larger; however, that would limit the $m$ available for spatial parallelism at the PE level, and also increase the tile size of $I$ and $Q$ that must be stored in the LLB.
\end{itemize}

\section{Case Study: TCM For Optimizing Architectures} \label{sec:case_study}
\insightbox{TCM enables lower-EDP accelerators and accurately informs design space explorations, while prior works misinform design space explorations because mapper limitations make their results unclear.}

To show the usefulness of TCM, we perform a design space exploration. This reflects a common goal for architecture designers; we would like to test many configurations of an accelerator and model how it will perform executing a benchmark.

In this exploration, we vary the size of the TPU global buffer between 64kB and 16MB while running the GPT3-6.7B~\cite{gpt3} workload on TPU-V4i-like with batch size 64 and sequence length 65,536. We also change the local buffer size to 16kB and, to yield more interesting mappings to explore, enable new dataplacement options, including letting any tensor bypass the global buffer and local buffer.

We compare TCM, Timeloop+Hint~\cite{timeloop}, and LOMA~\cite{zigzag,loma}. Timeloop without a hint is omitted, as its outputs are both far from optimal and too noisy to be used to inform a design space exploration. To ensure that the baselines return reasonable results, we give them $10\times$ the runtime of TCM.

\begin{figure}
    \centering
    \includegraphics[width=\linewidth]{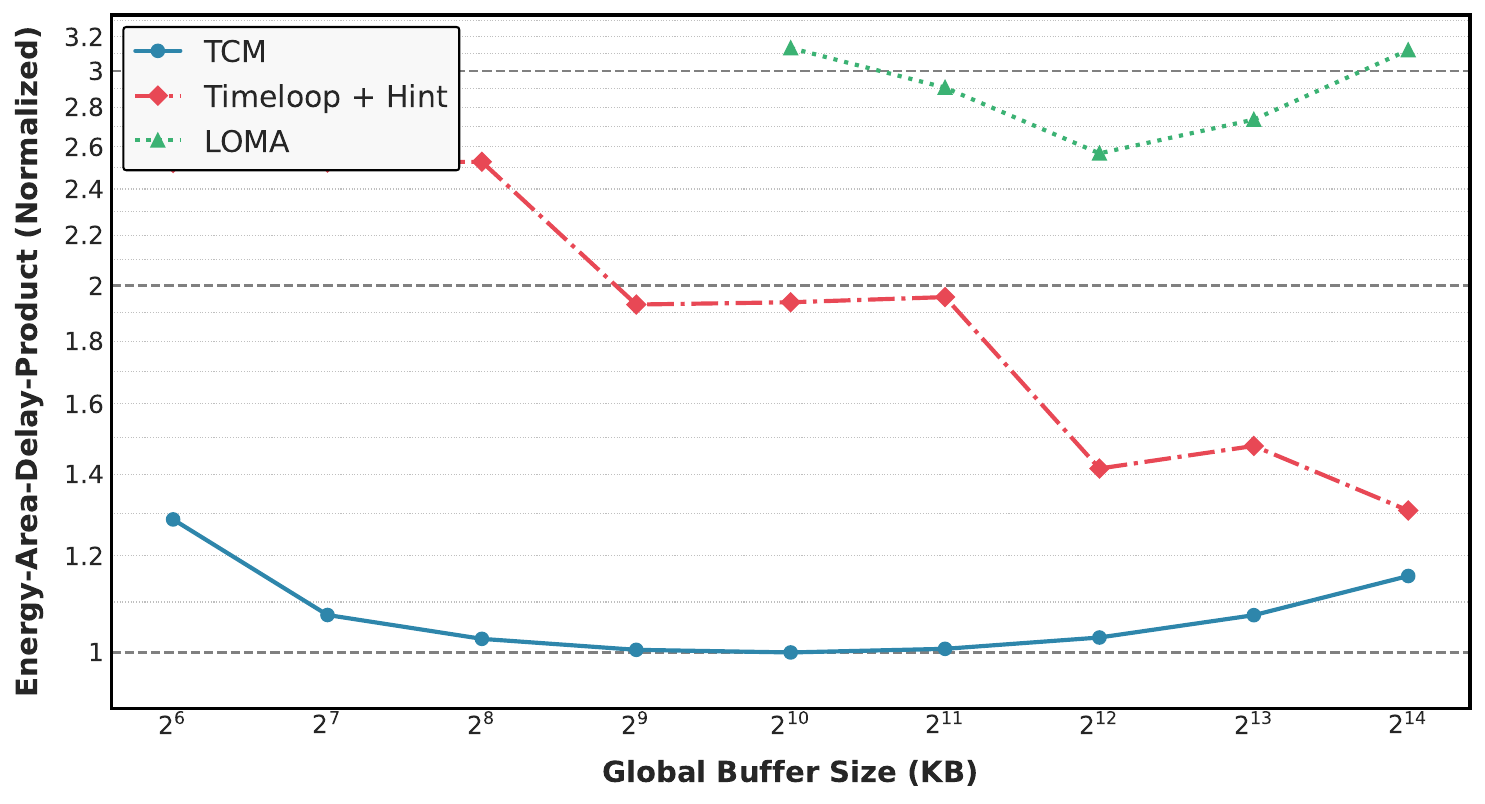}
    \caption{TCM accurately informs design space explorations, while prior works do not because mapper limitations hide the effect of architecture changes.   
    }
    \label{fig:sweep}
\end{figure}

Fig.~\ref{fig:sweep} shows the energy-area-delay-product (EADP) of the accelerator at each global buffer size. We can see that TCM produces a smooth curve, with a clear minimum in the range of 512kB-1MB.

The baselines, however, tell different and misleading stories. Timeloop+Hint reports that the best EADP can be found at a global 16MB global buffer size ($16\%$ off optimal) while LOMA reports that the best global buffer size is 4MB ($6\%$ off optimal).

Moreover, both baselines obscure the impact of this architecture decision because their results are a result of mapper limitations, not architecture changes. Timeloop+Hint steadily improves as global buffer size increases because it becomes easier to find valid mappings. Meanwhile, LOMA fails to find valid mappings for global buffer sizes less than 1MB and produces far-from-optimal mappings for other global buffer sizes. This is because, rather than performing a comprehensive search, LOMA forces full spatial utilization then optimizes one memory level at a time. This leads to the optimizer picking large tiles that overwhelm the global buffer.

\begin{figure}
    \centering
    \includegraphics[width=\linewidth]{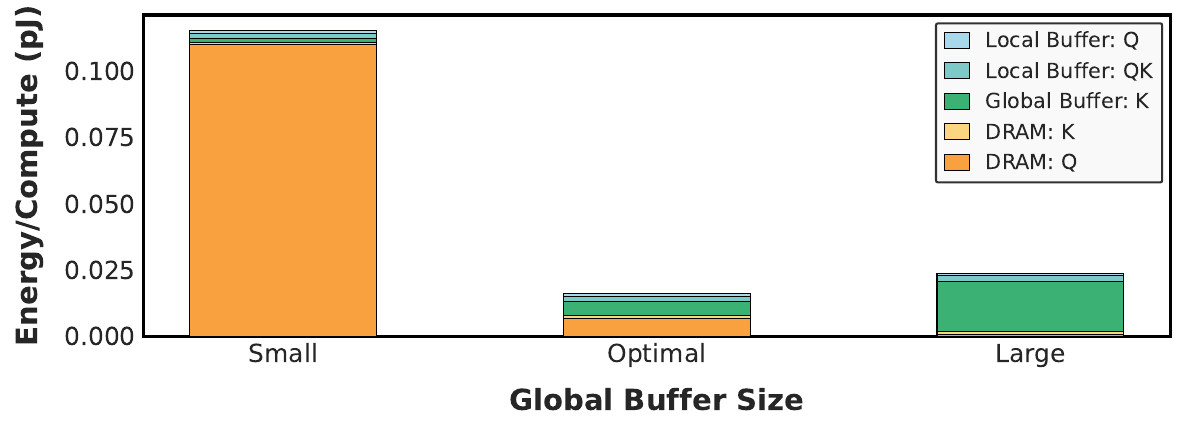}
    \caption{TCM balances the energy of multiple memory levels to yield low-energy accelerators. The optimal design uses a moderately-sized global buffer to reduce DRAM accesses while costing moderate global buffer access energy. To make differences clearer, we omit irreducible costs of MACs and writeback of the output tensor $QK$ to DRAM (which TCM does only once), noting that Fig.~\ref{fig:sweep} includes all components.
    }
    \label{fig:breakdown}
\end{figure}

Fig.~\ref{fig:breakdown} shows the energy breakdown of the accelerator at small (64kB), optimal (1MB), and large (16MB) global buffer sizes. To make differences clearer, we omit large unavoidable costs of performing MAC operations and writeback of the output tensor $QK$ to DRAM (which TCM does only once). Note that these costs dominate overall energy overall delta can be seen Fig.~\ref{fig:sweep}.

We can see different effects taking place at each global buffer size. For the small global buffer, TCM cannot capture all reuse at the GLB level and must tile, leading to tensor $Q$ being refetched many times from DRAM. Increasing global buffer size to the optimal level lets TCM capture additional reuse and reduce the amount of $Q$ tiling. However, the larger global buffer now has higher $K$ energy because larger memories consume more energy to access. As the global buffer size and access energy increase further, this trend continues, with higher global buffer $K$ energy and lower DRAM $Q$ energy. 

Notice that the optimal global buffer size is \emph{not} big enough to capture all reuse; rather it balances a large global buffer to reduce DRAM fetches and a small global buffer to reduce access energy. The sweet spot will change for different workloads, architectures, and technologies, so it is critical to have a mapper that can effectively identify the optimal tradeoff.

%% file: sections/7.conclusion.tex
\section{Conclusion}
In this paper, we have shown the important new concept of \emph{dataplacement}, which, alongside dataflow and tile shapes, provides a clear way to understand and analyze mappings. Defining and decoupling dataplacement lets us perform new mapping analyses, letting us identify and prune suboptimal choices and create a much more efficient curried model. We show that these improvements make it possible, for the first time, to fully explore mapspaces and find optimal mappings, yielding better accelerator energy, throughput, and latency.
